\documentclass[twocolumn,tighten]{aastex63}

\usepackage{color}

\usepackage{graphicx}	% Including figure files
\usepackage{amsmath}	% Advanced maths commands
\usepackage{amssymb}	% Extra maths symbols

\accepted{November 19, 2019}

\begin{document}

%%%%%%%%%%%%%%%%%%%%%%%%%%%%%%%%%%%%%%%%%%%%%%%%%%

%%%%% AUTHORS - PLACE YOUR OWN COMMANDS HERE %%%%%

% Please keep new commands to a minimum, and use \newcommand not \def to avoid
% overwriting existing commands. Example:
%\newcommand{\pcm}{\,cm$^{-2}$}	% per cm-squared

%%%%%%%%%%%%%%%%%%%%%%%%%%%%%%%%%%%%%%%%%%%%%%%%%%

%%%%%%%%%%%%%%%%%%% TITLE PAGE %%%%%%%%%%%%%%%%%%%

% Title of the paper, and the short title which is used in the headers.
% Keep the title short and informative.
\title[]{Exploring the outskirts of globular clusters: the peculiar kinematics of NGC3201}
%The peculiar kinematics in the outskirts of the globular cluster NGC3201}

% The list of authors, and the short list which is used in the headers.
% If you need two or more lines of authors, add an extra line using \newauthor
\author[0000-0002-0358-4502]{P. Bianchini} 
\affiliation{Observatoire Astronomique de Strasbourg, Universit\'{e} de Strasbourg, CNRS UMR7550, F-67000 Strasbourg, France} 
\author[0000-0002-3292-9709]{R. Ibata} 
\affiliation{Observatoire Astronomique de Strasbourg, Universit\'{e} de Strasbourg, CNRS UMR7550, F-67000 Strasbourg, France} 
\author[0000-0003-3180-9825]{B. Famaey} 
\affiliation{Observatoire Astronomique de Strasbourg, Universit\'{e} de Strasbourg, CNRS UMR7550, F-67000 Strasbourg, France} 

\correspondingauthor{Paolo Bianchini} 
\email{paolo.bianchini@astro.unistra.fr}
% These dates will be filled out by the publisher
%\date{Accepted XXX. Received YYY; in original form ZZZ}

% Enter the current year, for the copyright statements etc.

% Don't change these lines

% Abstract of the paper
\begin{abstract}
The outskirts of globular clusters (GCs) simultaneously retain crucial information about their formation mechanism and the properties of their host galaxy. Thanks to the advent of precision astrometry both their morphological and kinematic properties are now accessible. Here we present the first dynamical study of the outskirts of the retrograde GC NGC3201 until twice its Jacobi radius ($<100$ pc), using specifically-selected high-quality astrometric data from \textit{Gaia} DR2. We report the discovery of a stellar overdensity along the South-East/North-West direction that we identify as tidal tails. The GC is characterized globally by radial anisotropy and a hint of isotropy in the outer parts, with an excess of tangential orbits around the lobes corresponding to the tidal tails, in qualitative agreement with an $N$-body simulation. Moreover, we measure flat velocity dispersion profiles, reaching values of $3.5\pm0.9$~km~s$^{-1}$ until beyond the Jacobi radius. While tidal tails could contribute to such a flattening, this high velocity dispersion value is in disagreement with the expectation from the sole presence of potential escapers. To explain this puzzling observation, we discuss the possibility of an accreted origin of the GC, the presence of a dark matter halo -- leftover of its formation at high redshift -- and the possible effects of non-Newtonian dynamics. Our study uncovers a new path for the study of GC formation and of the properties of the Milky Way potential in the era of precision astrometry.  
\end{abstract}

% Select between one and six entries from the list of approved keywords.
% Don't make up new ones.

%%%%%%%%%%%%%%%%%%%%%%%%%%%%%%%%%%%%%%%%%%%%%%%%%%

%%%%%%%%%%%%%%%%% BODY OF PAPER %%%%%%%%%%%%%%%%%%

\section{Introduction}
%------------------------------------------------------------------%
\begin{table*}
\begin{center}
\caption{\textbf{Structural and orbital properties of NGC3201}}.
\begin{tabular}{lccccccccccc}
\hline\hline
d$_\odot$ & d$_\mathrm{GC}$ & pericentre & apocentre & $M$ &$\log T_\mathrm{rel}$ & \multicolumn{2}{c}{$r_h$} & \multicolumn{2}{c}{$r_t$} & \multicolumn{2}{c}{$r_j$}\\
kpc & kpc & kpc & kpc & M$_\odot$ & yr & pc & armin & pc & arcmin & pc & arcmin\\
 %$^{(1)}$& & & & & & & & & &\\
\hline
4.9  $^{(1)}$& 8.8  $^{(1)}$ & 8.5  $^{(2)}$ & 22-29  $^{(2)}$& $3.98\times10^5$  $^{(3)}$ & 9.27  $^{(1)}$& 4.41  $^{(1)}$ &  3.10 & 36.11 $^{(1)}$& 25.35 & 83.46 $^{(4)}$ & 58.58 \\

\hline

\end{tabular}
\tablecomments{Distance to the sun $d_\odot$, distance to the galactic centre $d_{GC}$, pericentre and apocentre, total mass $M$ from Wilson model fit, half-light relaxation time $T_\mathrm{rel}$, half-light radius $r_h$, tidal truncation radius $r_t$, Jacobi radius $r_j$. (1) \citet[][2010 edition]{Harris1996}, (2) \citet{Helmi2018}, (3) \citet{MLvdM2005}, (4) \citet{BalbinotGieles2018}.}
\label{tab:1}
\end{center}
\end{table*}
%------------------------------------------------------------------%

The origin of globular clusters is one of the open questions in modern astrophysics. Numerous pieces of evidence indicate that some of the Milky Way (MW) GCs formed in accreted dwarf galaxies (e.g. \citealp{Leaman2013,Massari2019}); yet it is still unclear whether they originated in the early universe within dark matter mini-halos or simply as gravitationally bound clouds (e.g. \citealp{Peebles1984,Penarrubia2017}). 

Important clues on the origin of GCs come from the study of their current properties. Despite the fact that their structure is strongly shaped by their $>10$ Gyr long internal processes (relaxation processes and stellar evolution) and by the gravitational interaction with their host galaxy, primordial long-lasting features imprinted at formation can survive. In particular, the outskirts of GCs, characterized by long relaxation times, are the unique environment to hunt for these primordial properties. Moreover, they represent the test bed to study the transition between internally-driven dynamical processes (e.g., internal relaxation, evaporation) and external processes driven by the host galaxy (e.g. tidal stripping, formation of tidal tails). Therefore, the study of the outskirts of GCs offers important clues on their long term evolution as well as on the properties of the MW potential. 

Observationally, a number of studies have shown that the outskirts of some GCs are characterized by extended stellar structures and tidal debris beyond their nominal tidal radius (\citealp{Grillmair1995, Leon2000, Chun2010, Kuzma2018}). Kinematically, a few line-of-sight velocity studies have indicated that some GCs exhibit flattened velocity dispersion profiles (\citealp{Drukier1998,Scarpa2007,Lane2010, DaCosta2012,Bellazzini2015}). Different -- but yet not conclusive -- explanations for these phenomena have been put forward: i) presence of energetically unbound stars still orbiting within the GCs (potential stellar escapers, \citealp{FukushigeHeggie2000,Kupper2010}), ii) presence of dark matter, relic of a primordial sub-halo in which GCs formed (e.g. \citealp{Ibata2013}), iii) modified Newtonian dynamics (e.g. \citealp{Hernandez2013}). A full morphological and dynamical characterization of the outer regions of GCs is therefore still needed to disentangle these scenarios.

With the recent advances in the field of precision astrometry led by \textit{Gaia} DR2 (\citealp{Brown2018,Helmi2018}), the study of \textit{both the kinematic and morphological properties} of (virtually all) MW GCs up to their outer regions is now possible (see e.g. \citealp{Bianchini2018}). In this Letter, we report the first dynamical study of a GC, NGC3201, from its intermediate regions until beyond its Jacobi radius ($\lesssim100$ pc) using \textit{Gaia} DR2 proper motions. In this work, we will refer to the tidal truncation radius $r_t$ as an approximation of the boundary of the GC obtained from fitting its surface brightness profile. The actual physical border of the cluster is identified by the Jacobi radius, $r_j$, that is the distance between the first Lagrangian point and the centre of the cluster, roughly representing the boundary radius inside of which the cluster is gravitationally dominant with respect to the Galaxy.

NGC3201, at a distance of 4.9 kpc (\citealp{Harris1996}), is characterized by a retrograde and eccentric orbit (\citealp{Helmi2018}), an intermediate relaxation time ($\log T_\mathrm{rel}/\mathrm{yr}=9.27$, \citealp{Zocchi2012}) suggesting a not very efficient internal dynamical evolution, and the presence of a few extra-tidal stars (\citealp{Kunder2014,Anguiano2016,Kundu2019}). We initially spotted the presence of the tidal tail around NGC3201 by searching close to the Galactic plane with the $\texttt{STREAMFINDER}$ algorithm \citep{Malhan2018}, and noticed that it appears to link up with the Gj\"oll stream \citep{Ibata2019}. These new stream maps will be presented in a forthcoming contribution (Ibata et al., in prep). These properties (reported in Table \ref{tab:1}) make NGC3201 an ideal target to pioneer studies on the external regions of GCs and to unveil their formation and evolution.

\section{Cluster members selection}
\label{sec:membership}

%---------------------------------------------------------------------------%
\begin{figure*}
\centering
\includegraphics[width=0.9\textwidth]{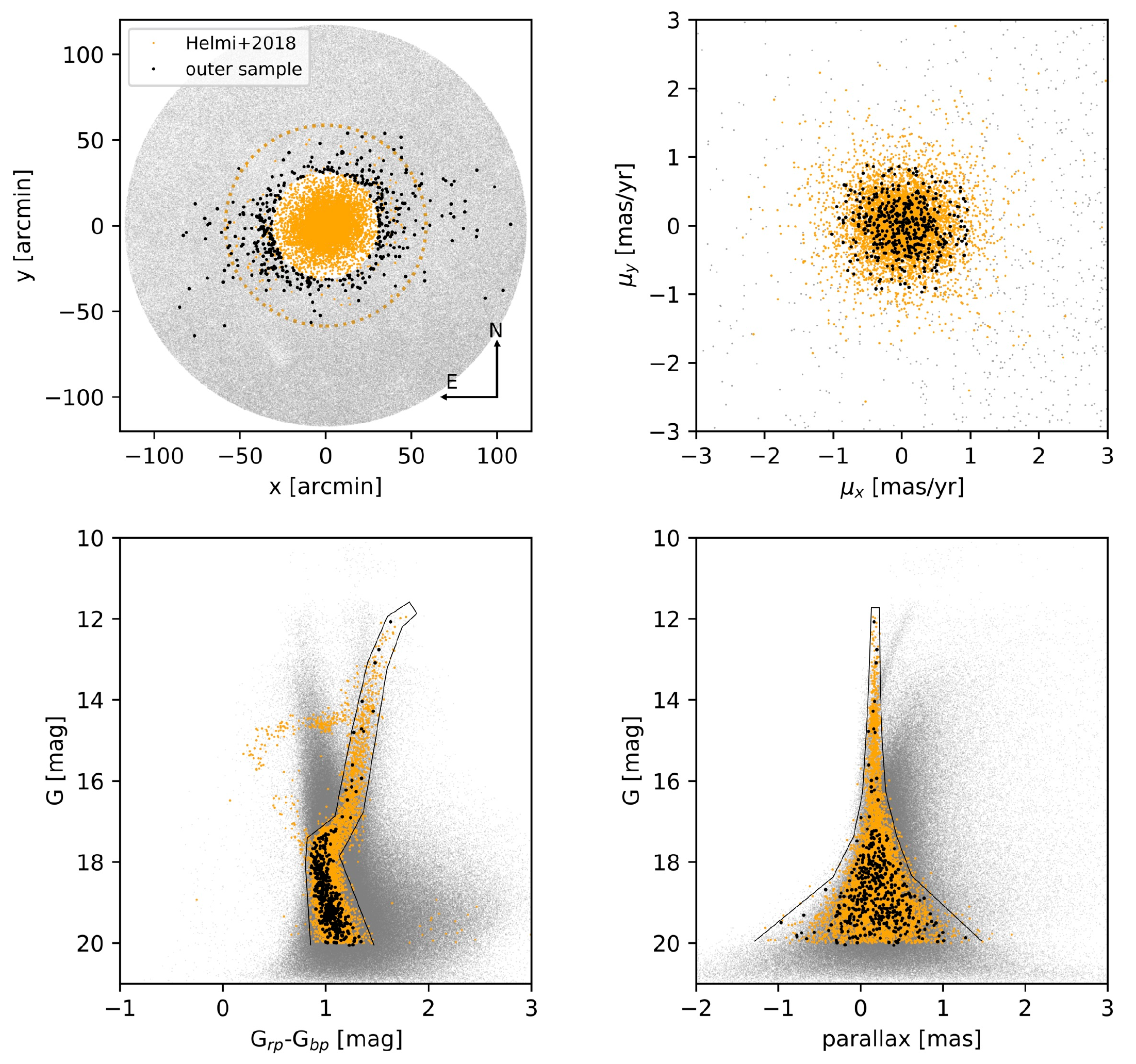}
\caption{NGC3201 member stars selection in the area extending to $2\,r_j$. The orange points are the inner sample member stars from \citet{Helmi2018}, the grey points all the stars in the $2\,r_j$ FoV,  and the black dots are the outer stars selected as members in this paper. The membership selection is based on the properties of the $\mu_x-\mu_y$ diagram (\textit{top right plot}), colour-magnitude diagram (\textit{bottom left plot}) and the parallax vs. $G$ magnitude diagram (\textit{bottom right plot}) of the inner sample. Only high-quality proper motion measurements suitable for dynamical analysis are included in the sample. \textit{Top left:} the FoV of NGC3201 clearly shows the presence of tidal tail structures in the outskirt of the cluster, extending beyond the Jacobi radius (dashed line).}
\label{fig:FoV}
\end{figure*}
%---------------------------------------------------------------------------%

To probe the dynamics of the outer regions of a GC we need to select high-quality proper motion measurements suitable for a kinematic analysis, and, at the same time, minimize background and foreground contaminations from likely non-cluster members. As a first step, we consider the member stars used in \citet{Helmi2018} and perform a series of quality cuts to select high-precision astrometric data and avoid measurements affected by crowding. These cuts, based on parameters provided in the \textit{Gaia} DR2 catalog and following the procedures illustrated in \citet{Lindegren2018} (see also \citealp{Vasiliev2018,Jindal2019}), include:
\begin{itemize}
\item $\texttt{astrometric\_gof\_al}<0.5$;
\item $\texttt{astrometric\_excess\_noise}<1$;
\item $\texttt{phot\_bp\_rp\_excess\_factor}$$<1.3$+$0.06\,(\texttt{bp-rp})^2$;
\item $\texttt{ruwe}<\texttt{ruwe}_{95}$, that is the 95-percentile of the renormalized unit weight error, $\texttt{ruwe}$ (see \url{https://www.cosmos.esa.int/web/gaia/dr2-known-issues} for a definition).
\end{itemize}
This brings the sample from a total of 19922 to 8222 stars within the Jacobi radius of the cluster ($r_j$=83.46 pc/58.58 arcmin, \citealp{BalbinotGieles2018}). To extend the sample beyond the Jacobi radius of the cluster we select an outer sample of stars within $10\, r_h<r<2\,r_j$, with $r_h=3.10$ arcmin, half-light radius of the cluster (\citealp{Harris1996}, 2010 edition), allowing an overlap with the inner sample. We apply the same high-quality proper motion criteria employed above and select the member stars based on proper motions, $G_{rp}-G_{bp}$ colour,  $G$ magnitude and parallax, using $\texttt{TOPCAT}$. This step is visualized in Figure \ref{fig:FoV}. The stars, in (ra,dec) and ($\mu_\mathrm{ra}$,$\mu_\mathrm{dec}$), are projected into a Cartesian coordinate system ($x,y$) and ($\mu_x,\mu_y$) using eq. 2 of \citet{Helmi2018}, with positive $x$- and $\mu_x$-axes pointing West. We subtract from the velocities the mean motion of NGC3201, ($\mu_x,\mu_y$)=$(-8.3344,-1.9895)$ mas yr$^{-1}$.

First, stars from the outer sample are selected to match the colour-magnitude properties of the inner sample and the parallax vs. $G$ magnitude diagram. Subsequently, only stars within 1 mas yr$^{-1}$ of the $\mu_x-\mu_y$ diagram are considered, giving a sample of 8271 stars. We also consider a looser cut in the velocity space that includes all stars within 2 mas yr$^{-1}$ (see Section \ref{sec:kin}). A clear overdensity is visible along the South-East/North-West direction in the FoV of the cluster (first panel Figure \ref{fig:FoV}), indicating the discovery of tidal tails structures around NGC3201. This observed feature is independent of the cut adopted in the velocity space. 

Since field contamination can be very critical in the outer part of the cluster, we follow a decontamination technique similar to the one outlined in \citet{Carballo-Bello2019}. We select a field sample ranging from $2\,r_j<r<2.5\,r_j$ and apply the same quality cuts as the ones applied to the members. For a given four-dimensional bin of magnitude, colour, $\mu_x$ and $\mu_y$, we calculate the surface number density of field stars and cluster stars $n_\mathrm{field}$ and $n_\mathrm{cluster}$ and we associate to each star a contamination factor $w=1- n_\mathrm{field}/n_\mathrm{cluster}$. We consider magnitude cells of 0.5 mag, colour cells of 3 mag and velocity cells of 2 mas yr$^{-1}$, in the intervals $8<G<21$ mag, $-1<G_{rp}-G_{bp}<4$ mag, $-4<\mu<4$ mas yr$^{-1}$. We further recalculate $w$ after shifting the starting position in steps of 1/10, 1/5 and 1/10 , in magnitude, colour and velocity, respectively, in order to minimize the low-number statistical fluctuation due to the bin selection.

Globally, our analysis shows that the contamination level expected in our sample is $\sim13\%$; this indicates that in the outskirts of the cluster, prior to decontamination, only approximately $\sim13$ stars (out of the 102 stars with radius R$>$50 arcmin) are expected to be contaminants. For each star we compute the average contamination factor $w$ and we exclude from our sample those stars with $w<0.9$. A total of 33 stars are discarded, further indicating a low contamination from field stars.\footnote{We checked different values of the contamination threshold, and the final results are not affected by this choice.} Our final clean sample consists of 8238 NGC3201 member stars with high-quality proper motion measurements, extending from $\sim r_h$ beyond $r_j$ and include the tidal tail structure discovered here.

\section{Kinematic analysis}
\label{sec:kin}
We decompose the proper motions into radial and tangential components ($\mu_r,\mu_t$) and we propagate the velocity uncertainties and the covariances. The kinematic profiles (mean proper motion profiles, velocity dispersion profiles and anisotropy profile) are computed using the likelihood function presented in \citet{Bianchini2018} (see their eqs. 2 and 3) with the addition of the covariance term (see e.g. \citealp{Sollima2019}. eq. 3). The sample is divided in annular bins and for each bin the likelihood is sampled using the Markov Chain Monte Carlo algorithm $\texttt{emcee}$ by \citet{Foreman-Mackey2013}.

A biased assessment of the proper motions errors can have a high impact on the velocity dispersion measurements, particularly in the outskirts of the cluster where dispersions are expected to be low. We minimized this problem by selecting only high-quality astrometric measurements. Moreover, as shown in \citet{Brown2018}, \textit{Gaia} DR2 proper motion errors are underestimated by a factor of $\sim10\%$ for stars with magnitude $G>16$ and of $\sim30\%$ for stars with $G<13$. We include this additional correction before performing the kinematic measurements.

Finally, we test the presence of magnitude-dependent biases. For this purpose, we calculate the kinematic profiles for stars in magnitude bins (red giant branch $G<17$, $17<G<18$, $18<G<19$, $G>19$). We find that all the profiles are consistent with each other, indicating that any magnitude-dependent systematics are within the error uncertainties. Note that the range of magnitude covered corresponds to a stellar mass range of $\sim0.6-0.8$~$M_\odot$\footnote{Stellar masses from \texttt{PARSEC} isochrones \citep{Marigo2017}.}, therefore the lack of differences in velocity dispersion also indicates that the system, at least in the outer part $>r_h$, is far from a state of partial energy equipartition. This is fully consistent with the theoretical expectation for the outer areas of GCs where two-body relaxation processes are expected to be less efficient in shaping the internal dynamics (\citealp{TrentivanderMarel2013, Bianchini2016b, Bianchini2018b}).

%---------------------------------------------------------------------------%
\begin{figure}
\centering
\includegraphics[width=0.5\textwidth]{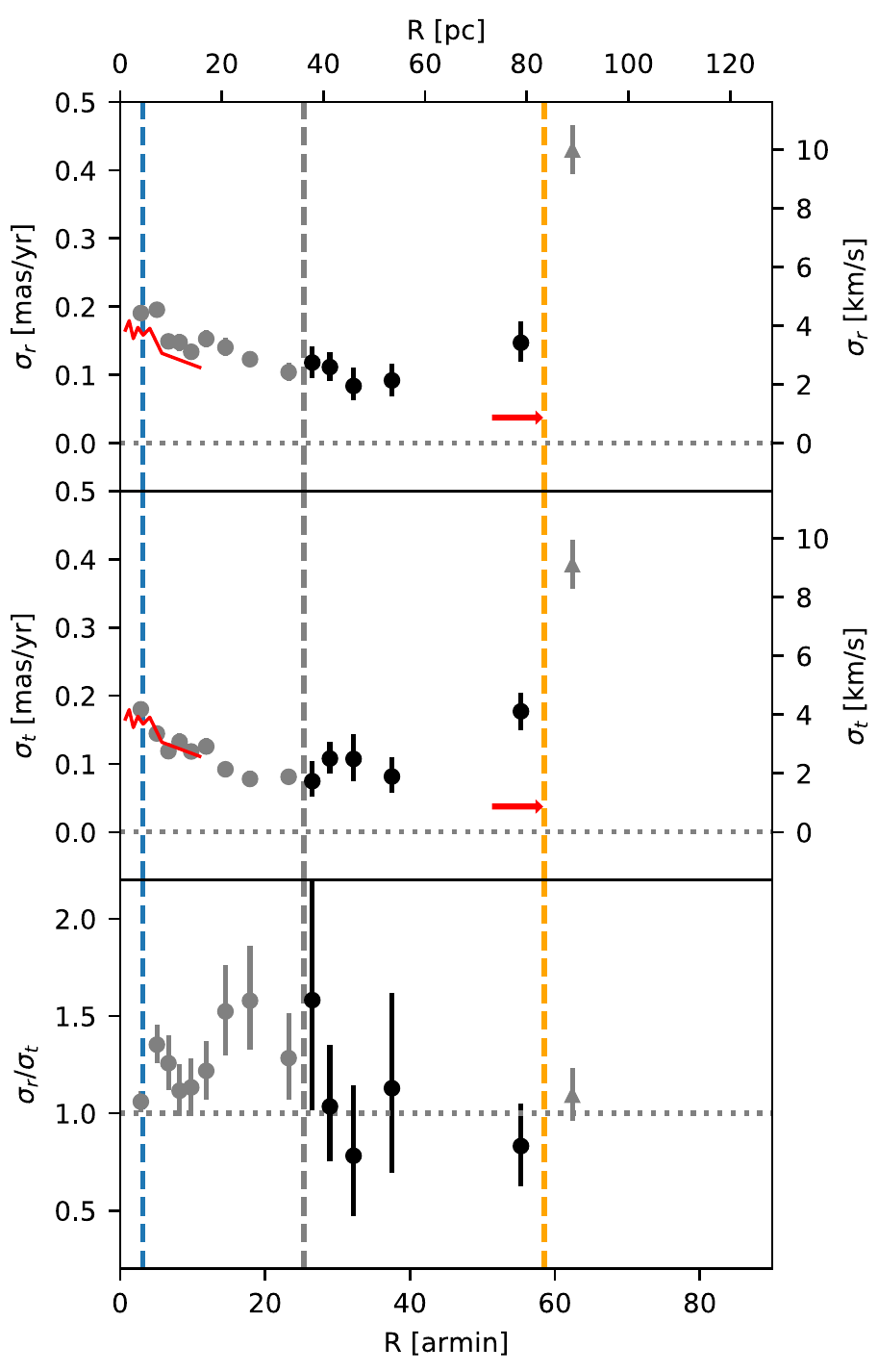}
\caption{\textit{From top to bottom:} Velocity dispersion profiles for the radial and tangential component of the proper motions $\sigma_r$ and $\sigma_t$, and velocity anisotropy profile. Grey circles refer to the region within the tidal radius, while the black circles refer to the outskirts of the cluster; the grey triangles correspond to the values obtained from a looser proper motion cut (see Section \ref{sec:membership}). The dashed lines are the half-light radius, tidal radius and Jacobi radius, in blue, grey, and orange, respectively. The red line is the line-of-sight velocity dispersion profile from \citet{Baumgardt2018}. The velocity dispersion profiles show an evident flattening in the outskirts and higher values than what is expected from the presence of potential escapers only (red arrow, \citealp{Claydon2017}). NGC3201 is radially anisotropic in the intermediate regions and isotropic in the outskirts.}
\label{fig:kin}
\end{figure}
%---------------------------------------------------------------------------%

In Figure \ref{fig:kin} we show the velocity dispersion profiles $\sigma_r$ and $\sigma_t$ and the anisotropy profile calculated as the ratio of the two. In order to increase the spatial resolution around the outskirts of the GC we employed the following binning: within the tidal radius (grey circles, $r_t=25.35$ arcmin; \citealp{Harris1996}, 2010 edition) each bin contains 823 stars; for the outer region (black circles) each bin contains 195 stars.

The inner parts of the velocity dispersion profiles are consistent with line-of-sight velocity measurements (\citealp{Baumgardt2018}; see red line in Fig. \ref{fig:kin}), while the outer parts, measured here for the first time, show a flattening around the Jacobi radius to values $\sim0.1-0.2$ mas yr$^{-1}$ ($\sim2.5-4$ km s$^{-1}$). We calculate a velocity dispersion at the Jacobi radius of $\sigma_r=0.14\pm0.04$ mas yr$^{-1}$ and $\sigma_t=0.16\pm0.04$ mas yr$^{-1}$ (with an average over the two components of $3.5\pm0.9$ km s$^{-1}$), defined as the velocity dispersion at R $>50$ arcmin, comprising 102 stars. The last point in each panel is derived using the looser selection cut in the velocity-velocity space (see Section \ref{sec:membership}) and presents a high value of velocity dispersion (0.5 mas yr$^{-1}$ / 11 km s$^{-1}$) that very likely indicates a stronger field contamination or the presence of unbound stars. We discuss the implication of the flattening of the velocity dispersion profiles in Section~\ref{sec:flattening}. 

The bottom panel of Figure \ref{fig:kin} shows that NGC3201 is characterized by radial anisotropy in the intermediate region, with a hint for a radial variation toward isotropy in the outskirts (although only marginally significant, also given the additional error systematics in Gaia DR2 of the order of $\sim0.07$ mas yr$^{-1}$, \citealp{Lindegren2018}). Globally, the cluster is characterized by a clear signature of radial anisotropy $\sigma_r/\sigma_t=1.25\pm0.04$ that can be taken as the result of internal relaxation processes, consistent with its intermediate relaxation condition (\citealp{Zocchi2012}). The trend of the anisotropy profile is in agreement with what found by \citet{Jindal2019} for a sample of MW GCs and suggests an isotropisation in the outskirts due to the effect of the Galactic tidal field that progressively strips stars on radial orbit (e.g. \citealp{BaumgardtMakino2003,Sollima2015,Tiongco2016,Zocchi2016,Bianchini2017b}; see also Figure \ref{fig:maps} in Section \ref{sec:sim}).

Finally, we note that the mean velocity profiles are consistent with the prediction obtained in \citet{Bianchini2018}: the radial component shows a perspective contraction due to the receding motion of the GC (eq. 4 of \citealp{Bianchini2018}; \citealp{vandeVen2006}) and the tangential component shows no significant sign of rotation within 0.05~mas~yr$^{-1}$/ 1 km s$^{-1}$. This result assures that data systematics are under control below the $\sim1$ km s$^{-1}$ level. 

\subsection{Comparison with a dynamical simulation}
\label{sec:sim}
%NGC3201 clearly shows the presence of tidal tails, therefore it is important to understand how they can affect the observations.
To understand the qualitative effects of tidal tails on GC kinematics, we analyze an $N$-body simulation that exhibits tidal tails in the final snapshot. The simulation was reported as DWL-MW10-evap in \citet{Bianchini2017b} and \citet{Miholics2016} and consists of N=50,000 initial particles evolved for 10 Gyr in a time dependent tidal field. The simulation was run with $\texttt{NBODY6TT}$ (\citealp{Renaud2015}) and includes a stellar mass function and stellar evolution.

\begin{figure*}
\centering
\includegraphics[width=0.68\textwidth]{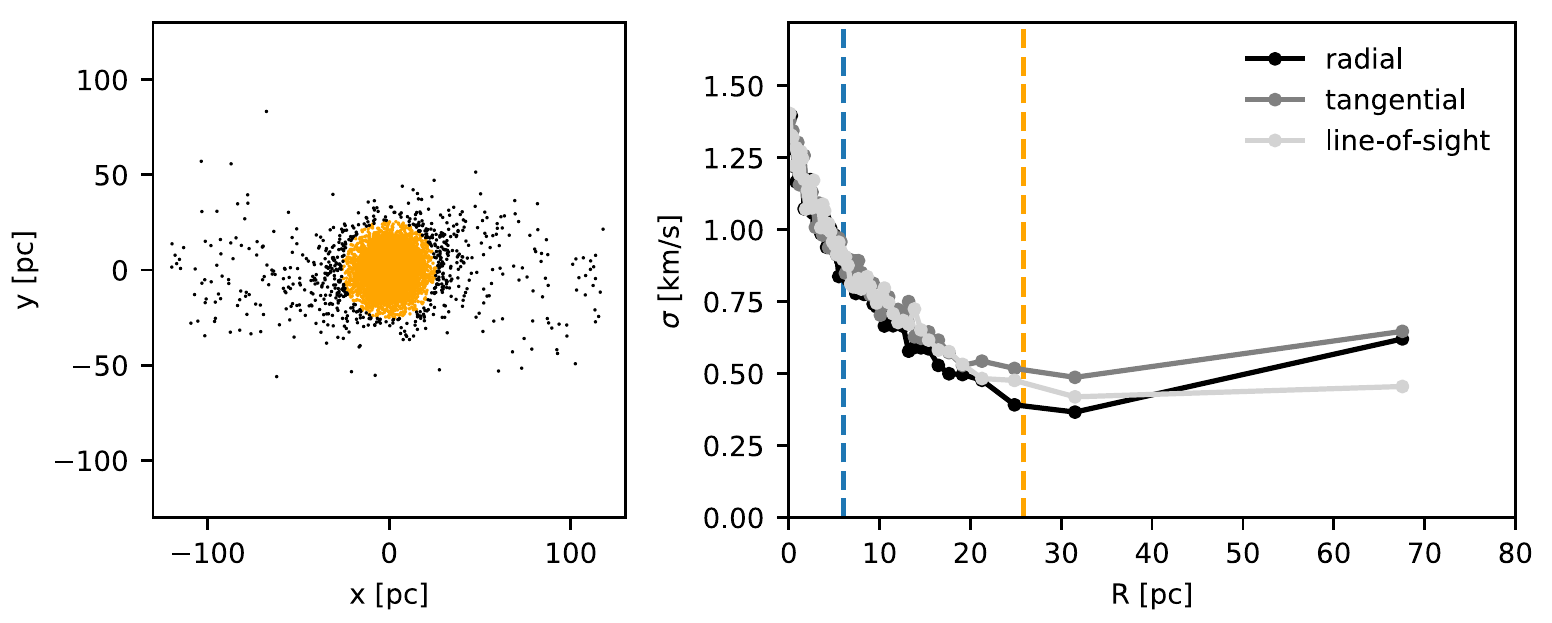}
\caption{Projected view of a snapshot at 10 Gyr of a $N$-body simulation forming tidal tails (see \citealp{Bianchini2017b,Miholics2016}). Orange points indicate stars inside the cluster's Jacobi radius. The right panel shows that the velocity dispersion profiles of the three velocity components flatten at radii larger than the Jacobi radius (orange dashed line). The blue dashed line is the half-light radius.}
\label{fig:sim_projections}
\end{figure*}

In Figure \ref{fig:sim_projections}, we show one projection of the simulation and the velocity dispersion profiles along the radial and tangential components of the proper motions and along the line-of-sight. The dispersion profiles show a behaviour qualitatively similar to the one we observed for NGC3201, indicating that tidal tails can contribute to the flattening of the velocity dispersion in the outskirts of the GC. Note, however, that in our observation the value of velocity dispersion at the Jacobi radius is similar to the one at the half-light radius, while for the simulation this is a factor $\sim2$ lower. This suggests that additional processes are contributing to the observed flattening of NGC3201 (see Section \ref{sec:flattening}); properly tailored simulations to NGC3201 are still needed for a conclusive comparison with data.

%---------------------------------------------------------------------------%
%---------------------------------------------------------------------------%

\begin{figure*}
\centering
\includegraphics[width=1.0\textwidth]{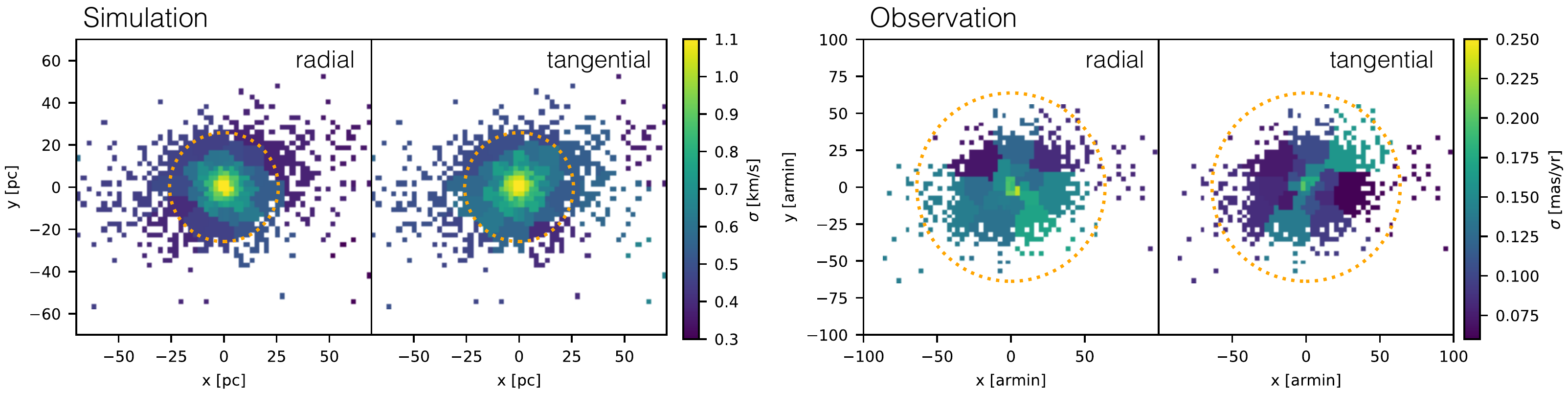}
\caption{Comparison between velocity dispersion maps of the simulation (\textit{left panel}) and the observation (\textit{right panel}). A higher tangential velocity dispersion is observed along the North-West/South-East direction, in correspondence with the tidal tail overdensities. This feature can be explained by stars in radial orbits preferentially escaping the clusters to form the tidal tails, leaving an overabundance of stars in tangential orbits in proximity of the Lagrangian points. The dashed lines represent the Jacobi radii.}
\label{fig:maps}
\end{figure*}

Furthermore, we compute the velocity dispersion maps, dividing the FoV in equal number spatial bins ($\approx300$ stars per bin), using the Voronoi binning algorithm (\citealp{CappellariCopin2003}). Figure \ref{fig:maps} shows the comparison between the simulated and observed maps. Both observation and simulation show a tangential velocity dispersion that is higher along the major axis of the system, connecting the points where the tidal tails begin. In contrast, the radial component of the velocity dispersion exhibits a feature that is perpendicular to that of the tangential component. These differences in the spatial distribution of velocity dispersions can be explained by the progressive stripping of stars on radial orbits by the Galactic tidal field, that will escape through the Lagrangian points and form the tidal tails. As a result, an excess of stars in tangential orbits are left around the Lagrangian points. This excess of tangential orbits allows us to identify the approximate location of the Lagrangian points, lying along the North-West/South-East direction.

%---------------------------------------------------------------------------%
%---------------------------------------------------------------------------%

\section{Flattening of the velocity dispersion profiles}
\label{sec:flattening}
The dynamics of the outskirts of GCs is known to be characterized by the presence of potential escapers, that is stars that are energetically unbound but still orbiting within the GC (e.g. \citealp{FukushigeHeggie2000,Kupper2010,Daniel2017}). The number of potential escapers is comparable to the number of bound stars around 0.5\,$r_j$ and dominates at larger radii (\citealp{Claydon2017}). As a result, they produce a flattening of the velocity dispersion profile. \citet{Claydon2017} provide a theoretical expectation of the velocity dispersion at the Jacobi surface given the GC mass, orbit, and the mass profile of the host galaxy. For NGC3201 this value is 0.87 km s$^{-1}$ as indicated in Figure \ref{fig:kin} with a red arrow. It is evident that our velocity dispersion measurement at the Jacobi radius $3.5\pm0.9$ km s$^{-1}$ is $\sim3$-sigma discrepant from the theoretical expectation.

This discrepancy can be taken as evidence of a peculiar dynamical history of NGC3201 or of some drawbacks due to the simplified framework adopted by \citet{Claydon2017} for the potential escapers calculation, which include many uncertain ingredients, such as the initial condition of the GC, its orbit and the Galactic potential. For example, if the cluster formed in a dwarf galaxy that later accreted onto the MW, it could have experienced a different tidal field than the one inferred by the current orbital properties. Therefore the calculation of potential escapers presented in \citet{Claydon2017} would not be valid. Interestingly, an accreted origin of the cluster is also suggested by its current retrograde orbit (in analogy to $\omega$ Cen, suspected to be the stripped nucleus of an accreted dwarf galaxy, e.g. \citealp{Freeman&Bland-Hawthorn2002,Ibata2019}). A similar scenario was recently suggested to interpret the morphology of the highly retrograde GD1 stellar stream, that is consistent with its progenitor having formed within a dark matter sub-halo, later accreted onto the MW (\citealp{Malhan2019b,Malhan2019}).

As an alternative, the presence of a dark matter halo embedding the cluster, as a relic of its formation environment at redshift $z>3$ (e.g. \citealp{Peebles1984,Ricotti2016}), could naturally explain the flattening of the velocity dispersion. Interestingly, \citet{Penarrubia2017} showed that the presence of a dark matter mini-halo can produce a gravitationally bound stellar envelope around a GC and a flattening of the velocity dispersion in the outskirts.

Finally, outside the Newtonian context, alternative theories of gravity such as MOND (e.g. \citealp{Famaey2012}) could be invoked to explain our observations. The typical acceleration in the outskirts of the cluster is of the order of $\sim3\,\sigma^2/R\approx0.05-0.1\,a_0$, placing it in the MOND regime ($a_0\sim1.2\times10^{-10}$ m s$^{-2}$, characteristic scale of MOND). MOND would predict a velocity dispersion in the outskirts of the order of $\sigma^4\approx\,G\,a_0\,M/\alpha^2$, with $\alpha$ the slope of the outer GC density profile. Given a value of $\alpha\approx-3.5$,\footnote{Calculated from the density profile of \citet{deBoer2019}.} the predicted velocity dispersion would be $\approx4-5$ km s$^{-1}$, remarkably similar to our observations. However, notice that the acceleration of the external field at the position of the cluster is of the order $\sim\,a_0$, larger than the typical acceleration of the GC itself. This external acceleration remains larger than the internal one even at apocenter. This implies that no flattening would be expected in MOND because of the external field effect (e.g. \citealp{Haghi2011,Derakhshani2014,Thomas2018}), apart from possible effects due to the varying tidal field along the highly eccentric orbit of the GC.

\section{Conclusions}
In this Letter, we presented the first kinematic analysis of the GC NGC3201, from the half-light radius to the outer regions beyond the Jacobi radius. We applied a strict selection to the \textit{Gaia} DR2 data in order to exclude non-member contamination and retain only high-precision astrometric measurements. Our sample reveals the presence of extended stellar structure beyond the nominal tidal radius. This tidal tails structure is aligned along the South-East/North-West direction and extends to $\lesssim 2\,r_j$ ($\lesssim100$ pc).

Thanks to the 2D kinematics, we measured an anisotropic distribution in velocity space, characterized by radial anisotropy in the intermediate regions and a hint of isotropy in the outer parts. These signatures are consistent with the combined effects of internal relaxation processes and ongoing action of the MW tidal field. Moreover, the velocity maps show an excess of tangential velocity dispersion around the lobes formed by the tidal tails, indicating that stars on radial orbits preferentially escape around the Lagrangian points. A comparison with a GC $N$-body simulation with tidal tails displays qualitatively similar features for the anisotropy configuration.

Our analysis reveals flattened velocity dispersion profiles starting from the tidal radius ($\approx40$ pc) until beyond the Jacobi radius, settling around values of $\sim2.5-4$ km s$^{-1}$, with an average velocity dispersion along the two components of $3.5\pm0.9$ km s$^{-1}$ at the Jacobi radius. This is in tension with what expected from the presence of potential escapers stars (\citealp{Claydon2017}) at a $\sim$3-sigma confidence level. However, note that we cannot exclude that small-scale systematics in the Gaia DR2 catalog could still produce an inflation of the velocity dispersion profiles of the order of $<1$ km s$^{-1}$. We propose that the observed discrepancy is due to a peculiar dynamical history of NGC3201, that could have formed in a dwarf galaxy later accreted onto the MW, as also suggested by its retrograde and eccentric orbit. Similarly, the presence of a dark matter halo around the cluster could naturally produce flattened velocity dispersion profiles. This would favour the idea that GCs primordially formed in dark matter mini-halo. Whether an alternative such as modified Newtonian dynamics could account for the observed discrepancy -- despite the external gravitational field dominates over the internal one -- remains to be studied in detail in further works.

Our work shows that GCs outskirts are the fundamental environment to unveil the formation properties of GCs, possibly detect the presence of dark matter and deviations from standard Newtonian dynamics. The extension of our analysis to more GCs, together with the augmented precision of future \textit{Gaia} data releases, the inclusion of line-of-sight data from future large surveys (e.g. WEAVE), and the comparison with tailored $N$-body simulations, will allow us to address these problems in greater detail.

\section*{Acknowledgements}
We thank the referee for the useful comments and suggestions. PB thanks Alice Zocchi, Anna Lisa Varri and Lorenzo Posti for several stimulating discussions. RI and BF acknowledge funding from the Agence Nationale de la Recherche (ANR project ANR-18-CE31-0006) and from the European Research Council (ERC) under the European UnionÕs Horizon 2020 research and innovation programme (grant agreement No. 834148). This work has made use of data from the European Space Agency (ESA)
mission {\it Gaia} (\url{https://www.cosmos.esa.int/gaia}), processed by
the {\it Gaia} Data Processing and Analysis Consortium (DPAC,
\url{https://www.cosmos.esa.int/web/gaia/dpac/consortium}). Funding
for the DPAC has been provided by national institutions, in particular
the institutions participating in the {\it Gaia} Multilateral Agreement.
This work made use of the PyGaia package provided by the Gaia Project Scientist Support Team and the Gaia Data Processing and Analysis Consortium (\url{https://github.com/agabrown/PyGaia}).

\bibliographystyle{aasjournal} % style aa.bst
\bibliography{biblio} % your references Yourfile.bib

%%%%%%%%%%%%%%%%%%%%%%%%%%%%%%%%%%%%%%%%%%%%%%%%%%

%%%%%%%%%%%%%%%%%%%% REFERENCES %%%%%%%%%%%%%%%%%%

% The best way to enter references is to use BibTeX:

%\bibliographystyle{mnras}
%\bibliography{example} % if your bibtex file is called example.bib

% Alternatively you could enter them by hand, like this:
% This method is tedious and prone to error if you have lots of references

%
% Don't change these lines

\end{document}